\documentclass[prl,twocolumn,nobibnotes,amsmath,amssymb,nofootinbib]{revtex4}

\usepackage{epsfig,mathptmx} 
 
\begin{document} 
 
\title{Periodic Field Emission from an Isolated Nano-Scale Electron Island }  
\draft 
\author{D. V. Scheible$^1$, C. Weiss$^2$, J. P. Kotthaus$^1$, and R. H. Blick$^3$} 
\address{$^1$Center for NanoScience and Fakult\"at f\"ur Physik der 
Ludwig-Maximilians-Universit\"at, Geschwister-Scholl-Platz 1, 
80539 M\"unchen, Germany\\ 
$^2$Institut f\"ur Physik, Carl von Ossietzky 
  Universit\"at, 26111 Oldenburg, Germany\\ 
$^3$Department of Electrical \& Computer Engineering, University of 
Wisconsin-Madison, 1415 Engineering Drive, Madison, Wisconsin 53706} 
\date{September 10, 2003} 
 
\begin{abstract} 
We observe field emission from an isolated nano-machined gold island. The 
island is able to mechanically oscillate between two facing electrodes,  
which provide recharging and detection of the emission current. We are able  
to trace and reproduce the transition from current flow through a rectangular  
tunneling barrier to the regime of field emission. A theoretical model via a  
master-equation reproduces the experimental data and shows deviation from the  
Fowler-Nordheim description due to the island's electric isolation.  
\end{abstract} 
 
\maketitle 
\narrowtext 
 
Field emission by microscopic tips has been a fundamental tool of experimental 
physics for decades. Deposited thorium at the end of a tungsten tip, acting as 
a field emitter, provided the first experimental visualization of single  
atoms \cite{muller_ZP1937}. Today, field emission from nano-scale emitters is 
subject to intense experimental and theoretical research \cite{bonard_PRL2002, 
  buldum_PRL2003, zheng_PRL2004, modi_N2003, pescini_AM2001, pirio_NT2002}. Bonard {\it et 
  al.} have studied the emission from individual carbon nanotubes 
\cite{bonard_PRL2002}, where deviation from the classical description of field 
emission by Fowler and Nordheim (FN) \cite{fowler_PRSA1928} is caused by 
geometrical effects and the particular electronic structure of the nanotubes 
\cite{buldum_PRL2003, zheng_PRL2004}. Purcell {\it et al.} have observed field 
emission from single nanotubes whilst the tubes were excited resonantly in one of their 
mechanical eigenmodes \cite{purcell_PRL2002}. In this Letter, we 
demonstrate field emission from an {\it isolated} nano-scale entity: a gold island is  
brought mechanically into an electric field configuration which provides 
the local field strength necessary for field emission. The island oscillates 
between the point of charge depletion toward one electrode, and the point at 
which it is recharged from a second electrode at the end of each cycle. Contrary to 
earlier observed deviation from the FN-formalism, the isolated nanomechanical 
pendulum shows new behavior already at low voltages. The 
fact that the emitter is isolated alters the FN-description to a 
behavior which becomes linear for large voltages.  
 
Hitherto conceived experiments of current spectroscopy in nano-scale 
electronic systems, such as laterally defined quantum dots 
\cite{holleitner_S2002, kouwenhoven_NATO1997}, mostly work in the regime of 
electrons tunneling through a barrier which is independent of the source-drain 
field. The same applies to nano-scale systems achieving electrical current  
transport across a structure with a mechanical degree of freedom 
\cite{erbe_PRL2001, park_N2000}. Mechanical displacement modulates the tunnel  
barrier and consequently regulates current transport  
\cite{gorelik_PRL1998} and enables the suppression of co-tunneling 
\cite{weiss_EPL1999}. The size of the island in the present device has been 
reduced six-fold compared to preceding work \cite{erbe_PRL2001}, thus 
increasing the electric field strength. As a consequence, the device undergoes the 
transition into the response of field emission and enhances strongly its net 
current up to several nano-Amp\`eres. In our setup field emission is 
controlled by both the voltage bias and the mechanical oscillations of the emitter.  
     
Operation of nano-electromechanical systems (NEMS), which have become an integral part of 
experimental mesoscopic physics \cite{blick_RMP2002, roukes_PW2001}, is 
predominantly carried out by the Lorentz force caused by an AC current in a 
perpendicular magnetic field. This magnetomotive drive requires high magnetic 
field densities (up to 15 Teslas), and consequently elaborate cryogenic 
cooling. In contrast to that, we drive our nanomechanical  
resonator by a combination of the capacitive force and the Coulomb force onto 
the excess charge being present on the shuttle island \cite{weiss_EPL1999, 
  scheible_APL2004}. The latter is strongly enhanced in the high current regime 
via field emission and therefore eases excitation. This mechanical excitation 
results in an oscillating shuttle between two facing gates, which provide the electric 
field and allow charging and discharging of the shuttle, when deflected toward  
the respective gate \cite{tuominen_PRL1999}. 
 
Our experimental setup consists of a nano-machined cantilever made from 
silicon-on-insulator material. At the tip of a freely suspended silicon 
cantilever of about 1 micron length we deposited an isolated gold island with dimensions 
$80 \times 80 \times 50$~nm$^3$. This resembles a small clamped bell. Two gates A 
and B face the grounded cantilever C [see Fig.~1(a)], and the island I 
oscillates between source S and drain D. AC excitation is applied to gate A, 
whereas an additional DC bias is imposed via source S. The resulting net 
current $I_{\rm D}$ is detected at drain D, and recorded versus AC excitation 
frequency $f$ with the DC bias $V$ as the parameter. We have presented the 
manufacturing process of such devices elsewhere \cite{scheible_NJP2002}.   
\begin{figure}[h]
 \epsfig{scale=1.0,width=0.9\linewidth,file=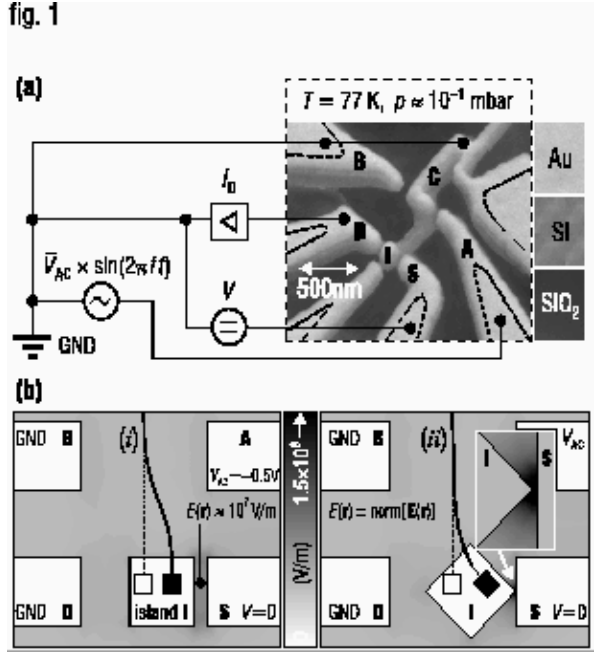} 
\caption{(a) SEM micrograph and experimental setup of the device: the electron 
  shuttle consists of a gold island I situated at the end of a nano-machined silicon 
  cantilever. Free suspension outside the area marked by the dotted lines is 
  $\sim 200$~nm above the sacrificial layer (SiO$_2$). The island oscillates between source S 
  and drain D, where the current $I_{\rm D}$ is detected via a current 
  amplifier. AC excitation is applied at frequency $f$ at gate A with a voltage amplitude 
  of $\bar V_{\rm AC} \approx 500$~mV. The DC bias $V$ is superimposed on 
  gate S. (b) Finite element solution of Maxwell's equations yielding the local 
  electric field strength $|{\bf E}({\bf r})|$, ranging from zero (white) to 
  1.0$\times$10$^8$~V/m (black) for an instantaneous external bias of $V_{\rm 
  AC}=-500$~mV at gate A, and a neutral island charge ($n_{\rm 
  I}=0$). In ({\it i}) the shuttle is deflected parallel toward drain, whereas 
  in ({\it ii}) the same deflection of the center of mass is assumed with an 
  edge facing the gate. The inset magnifies the facing tip with the absolute 
  field maximum of $6.9 \times 10^9$~V/m. This explains the manifestation of 
  field emission only for specific modes of excitation.}  
\end{figure}

As the bow-like bilayer system (Si/Au) strongly changes its dynamical 
response upon cooling the device, the AC excitation power $P$ has to 
be increased in order to maintain a stable current level through the 
nano-mechanical shuttle. Whereas at room temperature a reasonably low power in 
the range of $P=-30\ldots -10$~dBm suffices for operation 
\cite{scheible_JAP2004, scheible_APL2004}, we apply an incident AC 
power of $P=+8$~dBm at a device temperature of 77~K, at which all experiments 
were conducted. The resulting voltage drop at the NEMS itself lies between 
$V_0=\sqrt{PZ_0}$ for the system possessing a matched impedance 
$Z_0=50\Omega$, and $2V_0$ for infinite impedance. As we estimate the actual 
impedance above $1$~k$\Omega$ and power losses of the setup to 3~dB, the AC 
voltage amplitude will be roughly $\bar V_{\rm AC} \approx 500$~mV. This 
amplitude allows the NEMS to establish the transition from tunneling to field 
emission. The small gate-island distance of some tens of nanometers and local 
microscopic surface roughness, induced by the dry reactive ion  
etching \cite{pescini_AM2001}, support the manifestation of field emission. 
 
Single electron devices with a mechanical degree of freedom are well  
modelled by a master equation, describing the time dependence of the number of 
electrons on the oscillating island \cite{gorelik_PRL1998, weiss_EPL1999}: 
\begin{eqnarray} \label{eq:MasterEq} 
  \dot p(m,t) & = & - 
  \left[\Gamma^{+}_{\rm L}(m,t)+\Gamma^{+}_{\rm R}(m,t) 
  \right]p(m,t) \\ 
  & & - \left[ \Gamma^{-}_{\rm L}(m,t)+ 
  \Gamma^{-}_{\rm R}(m,t)\right]p(m,t) \nonumber\nopagebreak\\ 
  & & + \left[\Gamma^{+}_{\rm L}(m\!-\!1,t)+\Gamma^{+}_{\rm 
  R}(m\!-\!1,t)\right]p(m\!-1\!,t)  \nonumber\nopagebreak\\ 
  & & + \left[\Gamma^{-}_{\rm L}(m\!+\!1,t)+\Gamma^{-}_{\rm 
  R}(m\!+\!1,t)\right]p(m\!+\!1,t), \nonumber 
\end{eqnarray} 
where $\Gamma$ are the transition rates and $p$ the probability to 
find $m$ additional electrons on the island at time $t$. For devices in which 
charge current is established by tunneling, the golden rule transition rates 
are of the form~\cite{grabert_SCT1992}  
\begin{equation} 
\label{eq:gamma} 
 \Gamma_{\rm t} = \frac 1 {e^2R} \, \frac {\Delta E}{1-\exp\left(-\Delta 
 E/k_{\rm B}T\right)} \, . 
\end{equation} 
This approach was already successfully applied to explain earlier  
experiments~\cite{erbe_PRL2001} and also reproduces the response of the  
present device when still in the tunneling regime (see Fig.~2). Since the 
tunneling rate and therefore the resistance depends exponentially on the distance 
between gate and island, charge transport takes place only when the 
island is deflected toward one of the electrodes (co-tunneling can therefore 
be neglected). In the experimental regime where the field emission behavior is not yet 
visible, the tunneling time~$\tau$ is large compared to the effective contact 
time~$\propto 1/f$ \cite{weiss_EPL1999}: although the applied driving voltage 
acts as a rather large gate voltage on the island, the effective number $n$ of 
electrons transferred per period is of the order of $0<n<1$. 

 \begin{figure}[h] 
 \epsfig{scale=1.0,width=0.9\linewidth,file=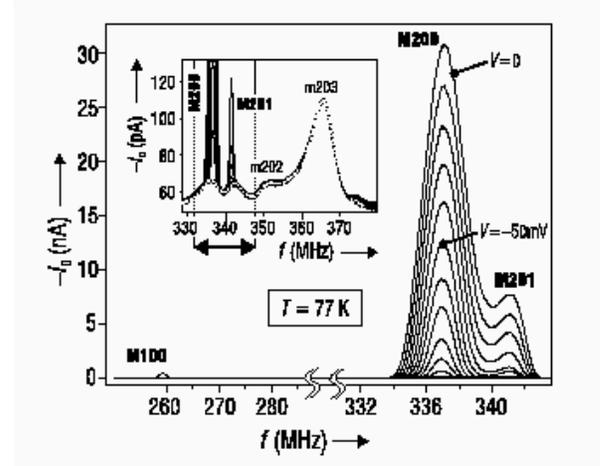} 
\caption{Spectral drain current $-I_{\rm D}$ vs excitation frequency $f$ for 
  bands in which field emission occurs first (DC bias $V$ ranges from 0 to 
  $-200$~mV). These three modes are denoted by M100, M200, and M201. The inset 
  shows the transition from tunneling to field emission in a magnified 
  current scale. The current differs roughly by two orders of magnitude. The 
  dotted line is a tunneling fit for a set of superimposed resonances, 
  according to the standard theory of a single electron shuttle of equations 
  (1) and (2). The arrow corresponds to the respective scope given in the main figure.} 
\end{figure}

For higher driving voltages (and thus larger electric fields between island and 
electrode), field emission between island and source electrode leads 
to a strongly positively charged island approaching the drain electrode. This 
results in a current which is several orders of magnitude larger. Applying a 
DC source voltage~$V$ will either increase the field (for positive voltages) or 
decrease the field (for negative voltages). In a single tunneling event, the 
tunneling probability decreases exponentially with distance $x$ as 
$\exp(-x/\lambda)$ for comparatively low electric fields. This can be seen in the 
tunneling resistance $R$ of Eq.~(2). For field emission this exponential 
decrease is replaced \cite{fluegge_PQM1994} by a factor of  
\begin{equation} 
\label{eq:trans} 
D = \exp\left[-\frac A{\cal E}\right],\quad A>0, 
\end{equation} 
where the electric field~${\cal E}$ in our case can be taken to be 
proportional to the voltage difference~$V+V_0$ between source electrode and 
island. The parameter $A$ accounts for the material specific emission 
behavior.  
 
However, even for a stationary pendulum, emission eventually stops as 
the island gets more and more positively charged. The total number of 
electrons will scale roughly linear \cite{gorelik_PRL1998, weiss_EPL1999} with the 
applied voltage. Thus, we can assume the number of 
electrons on the island~$n_{\rm I}$ after field emission (which only takes 
place for $V+V_0 > 0$) to be proportional to the product of the applied 
voltage and the transition probability~(\ref{eq:trans}), i.e. 
\begin{equation} 
\label{eq:cw} 
n_{\rm I} \propto  \left(V+V_0\right)\exp\left[-\frac B{V+V_0}\right]\;,\quad 
V+V_0>0. 
\end{equation} 
Geometrical details of the shuttle, as well as the material constant $A$, are 
contained in the parameter $B>0$. We describe the number of electrons on the  
island after field emission with Eq.~(\ref{eq:cw}) and the tunneling at the 
other electrode again numerically with equations (1) and (2). This results in a  
current proportional to the number of electrons on the island given by  
\begin{equation} 
  I = \left\{ 
\begin{array}{ccr} 
I_0\frac{V+V_0}{V_0}\exp\left[-\frac B{V+V_0}\right] &:& V+V_0>0 \\ 
0&:& V+V_0\le0 
\end{array}\,. 
 \right. 
\end{equation} 
This equation differs remarkably from ordinary field emission, since the 
emission takes place from an isolated entity: the $I$-$V$-characteristics 
lead to a linear dependence on the voltage $V$ for $V \gg V_0$, rather than the 
quadratic behavior predicted by FN-tunneling \cite{fowler_PRSA1928, 
  fluegge_PQM1994}. 
 
Experimental current traces are plotted in Fig.~2. In the main figure only 
three modes are visible in the high-current scale ranging up to 30~nA. These 
modes, denoted with a capital 'M' (M100, M200, and M201), show field 
emission. Magnification of the current axis reveals a second set of resonances 
which develop a much smaller absolute peak current of the order of only 50~pA 
(e.g. m202 and m203). This latter data can be fitted well by a set of 
superimposed resonances, assuming the shuttle transport to lie in the 
pure tunneling regime of Eq.~(2). Hence their denotation by a lower-case 
'm'. Furthermore, the response in the tunneling regime does not show 
dependence on the dc bias $V$, as already observed and analyzed earlier 
\cite{erbe_PRL2001}.   
 
If a mode sustains suitable mechanical response, i.e. island deflection toward 
gates is in a way that surface roughness or device edges result in a 
sufficiently high electric field strength [see mode ({\it ii}) in Fig.~1(b)], transition from 
tunneling current to the field emission current occurs. As shown in the inset of Fig.~2, 
some modes ignite (M200, M201) at particular frequencies, based on their 
respective tunneling current resonance. Adjacent modes however, such as m202 
and m203, do not show field emission transport for the given DC bias. In order to unambiguously 
attribute the observed behavior to field emission, we have applied a DC bias $V$ in the 
range of $-$500~mV$\ldots+$500~mV in addition to the AC power at $P=+$8~dBm. All 
modes M$j$ could be entirely detuned by a sufficiently negative DC bias $V$. With a 
superimposed DC bias of $V=-200$~mV and below field emission is suppressed, and the device 
is led back entirely into the tunneling regime. For an increasing voltage $V$ 
however (up to $+500$~mV), the current of the shown modes continues to rise in 
the discussed manner. Furthermore, more and more modes ignite and develop field emission 
response, as a positive bias increases the local field strength of any modal 
configuration of the island. The field emission offset $V_0$ here is characteristic for each 
mode M$j$. The observed peak current $I_{\rm D}^{\rm peak}$ scales linearly with the applied bias $V$ 
for $V \gg V_0$ as given by Eq.~(5). Figure 3 shows good accordance of the experimental 
values with this analytic fit, and gives evidence for the emission from the 
isolated shuttle island. 
  \begin{figure}[ht] 
 \epsfig{scale=1.0,width=0.9\linewidth,file=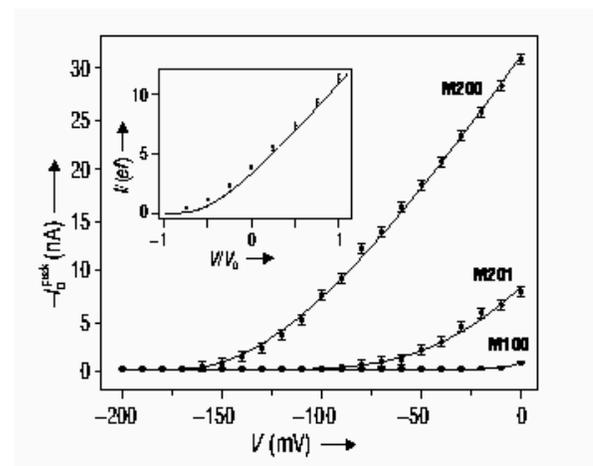} 
\caption{Peak current $I_{\rm D}^{\rm peak}$ vs DC bias voltage $V$ 
  for the first three modes which show transition to field emission. Error 
  bars correspond to maximum variation during the averaged recording of the 
  spectral current. The data have been fitted according to Eq.~(5). Inset: 
  If~$n_{\rm I}$ electrons are on the island after field emission, a 
  Monte-Carlo integration of the tunneling at the other electrode (points) 
  yields a current $I= g n_{\rm I} e f$ with $g < 1$ (solid line) and thus justifying the 
  analytic curve~Eq.~(5). The latter describes well the experimental results.} 
\end{figure}

Although at zero DC bias ($V=0$) a couple of modes show field emission, the majority of 
the dynamic response remains well within the tunneling regime. This is caused by the 
particular deflection of the island for each mode. We have modelled the 
magnitude of the electric field strength $E({\bf r})\equiv {\rm norm}[{\bf E}({\bf r})]$ via 
a finite element simulation \cite{FEMLAB} for the two most prominent cases: 
either the shuttle island faces a gate with the boundaries aligned parallel to 
each other, or an edge of the island --- respectively any other tip of the 
rough surface --- causes field enhancement. Both situations are drawn in 
Fig.~1(b), and the simulation shows an increase of the maximum field strength 
up to 6.9$\times$10$^{9}$ V/m, which suffices for field emission 
\cite{fluegge_PQM1994}. We have to stress that this is solely induced by a 
changed mode, keeping constant the deflection of the center of mass and the 
external bias. The finite element calculation assumes perfectly smooth surfaces rather than 
surface roughness. For the latter case, as it is in the experiment, the field 
strength is further increased.  
 
The nano-scale configuration of our device allowed the manifestation 
of field emission already at externally applied voltages below 1 Volt. The emission 
takes place from the isolated island toward one gate and a periodic 
oscillation allows recharging and hence quasi-continuous operation. Theoretical 
calculations, based on this model, well comply with the experimentally 
obtained data. Deviation from the classical FN-description of field emission 
can be attributed to the electrical isolation of the emitter. Excitation via 
the large excess charge on the shuttle was demonstrated and the manifestation 
of field emission in our NEMS yields a large current enhancement by a factor 
of 10$^2\ldots$10$^3$. We consider the combination of the environmentally 
sensitive field emission with the mechanical degree of freedom of a NEMS very 
promising for the interaction of the device with bio-molecules 
\cite{rief_S1997} and traces of specific chemicals \cite{modi_N2003}. Furthermore, application 
as a low-power loss radio-frequency filter appears feasible \cite{nguyen_IEEETMTT1999}. 
 
Authors would like to thank W. Zwerger and S. Manus for stimulating discussions, 
and A. Kriele and W. Kurpas for expert technical help. DVS gratefully acknowledges 
financial support by the Deutsche Forschungsgemeinschaft (DFG) under grant 
Bl-487 and the University of Wisconsin-Madison.

\end{document}